\documentclass[a4paper,english]{elsarticle}
\usepackage[T1]{fontenc}
\usepackage[latin9]{inputenc}
\usepackage{varioref}
\usepackage{units}
\usepackage{amsmath}
\usepackage{amssymb}
\usepackage{graphicx}
\usepackage{esint}
\usepackage{caption}
\usepackage{subcaption}

\makeatletter

\pdfpageheight\paperheight
\pdfpagewidth\paperwidth

\journal{Elsevier}
\usepackage{upgreek}
\usepackage[bookmarks,bookmarksopen,bookmarksdepth=6]{hyperref}
\usepackage{cases}
\usepackage{url}
\usepackage{lineno}

\usepackage{a4wide}

\makeatother

\usepackage{babel}
\begin{document}

\title{Study of the breakdown voltage of SiPMs}

\author[]{V.~Chmill$^{a, b}$}
\author[]{E.~Garutti$^{a}$}
\author[]{R.~Klanner$^{a,}$ \corref{cor1}}
\author[]{M.~Nitschke$^{a}$}
\author[]{and J.~Schwandt$^{a}$}

\cortext[cor1]{Corresponding author, Email: Robert.Klanner@desy.de,
  Tel. +49 40 8998 2558.}
\address{$^a$ Institute for Experimental Physics, University of Hamburg,
 \\Luruper Chaussee 149, D\,22761, Hamburg, Germany.}
\address{$^b$ Samara State Aerospace University,
 \\Moskovskoe Shosse 34, Samara 443086, Russian Federation.}


\begin{abstract}


 The breakdown behaviour of SiPMs (Silicon PhotoMultiplier) with pixel sizes of 15$\times $15, 25$\times $25, 50$\times $50, and 100$\times $100\,$\upmu $m$^2$, manufactured by KETEK, has been investigated.
 From the current-voltage characteristics measured with and without illumination by LED light of 470\,nm wavelength, the \textit{current-breakdown voltage}, $V_I$, and from linear fits of the voltage dependence of the SiPM gain, measured by recording pulse-area spectra, the \textit{gain-breakdown voltage}, $V_G$, have been obtained.
 The voltage dependence of the Geiger-breakdown probability was determined from the fraction of zero photoelectron events with LED illumination.
 By comparing the results to a model calculation, the \textit{photodetection-breakdown voltage}, $V_{PD}$, has been determined.
 Within experimental uncertainties, $V_I$ and $V_{PD}$ are equal and independent of pixel size.
 For $V_G$, a dependence on pixel size is observed.
 The difference $V_I - V_G$ is about 1\,V for the SiPM with 15\,$\upmu $m pixels, decreases with pixel size and is compatible with zero for the SiPM with 100\,$\upmu $m pixels.

\end{abstract}


\maketitle
 \tableofcontents
 \pagenumbering{arabic}

\section{Introduction}
 \label{sect:Introduction}
 Silicon photomultipliers (SiPMs), pixel arrays of avalanche photodiodes operated above the breakdown voltage, are becoming the photodetectors of choice for many applications.
 Thus simple and reliable methods of SiPM characterization with high throughput are of significant interest.
 Two parameters, which are relevant for the SiPM user, are the breakdown voltage, $V_{bd}$, and the Gain, $G$.
 Usually it is assumed that for an applied voltage $V$, the gain $G(V) \approx C_{pix}\cdot (V - V_{bd})/q_0$, with the single pixel capacitance, $C_{pix}$, and the elementary charge, $q_0$.
 In this contribution three different methods are used to determine the breakdown voltages for SiPMs with four different pixel sizes fabricated by KETEK\,\cite{KETEK}.
 We find that the different methods result in different values of the breakdown voltage.
 These observations are also relevant for the understanding and simulation of the breakdown mechanism and the response of SiPMs.

 \section{Sensors investigated}
  \label{sect:Experiment}
\begin{figure}[!ht]
   \centering
   \begin{subfigure}[a]{0.4\textwidth}
    \vspace{1.5cm}    
    \includegraphics[width=\textwidth]{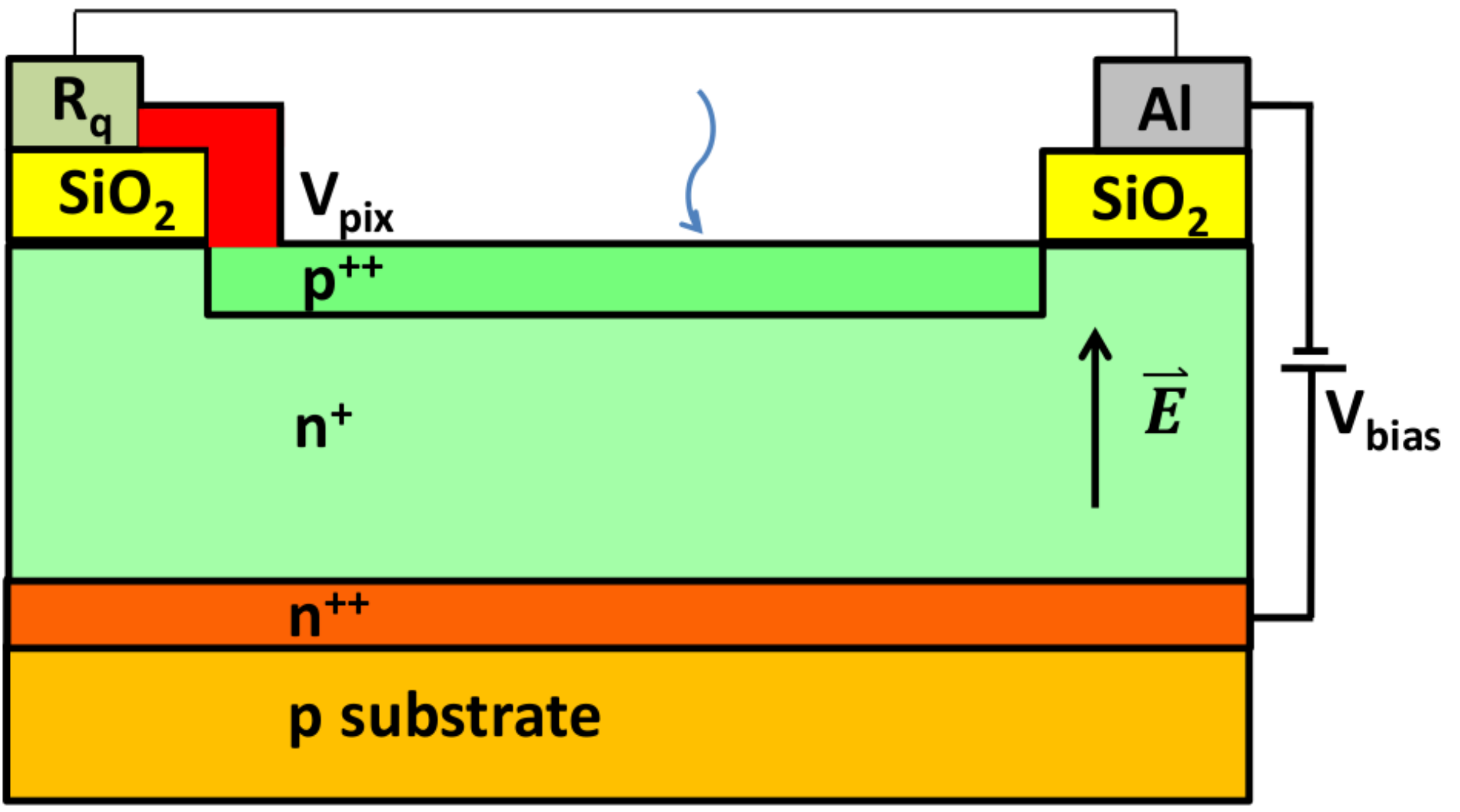}
    \caption{ }
   \end{subfigure}%
    ~
   \begin{subfigure}[a]{0.3\textwidth}
    \includegraphics[width=5cm]{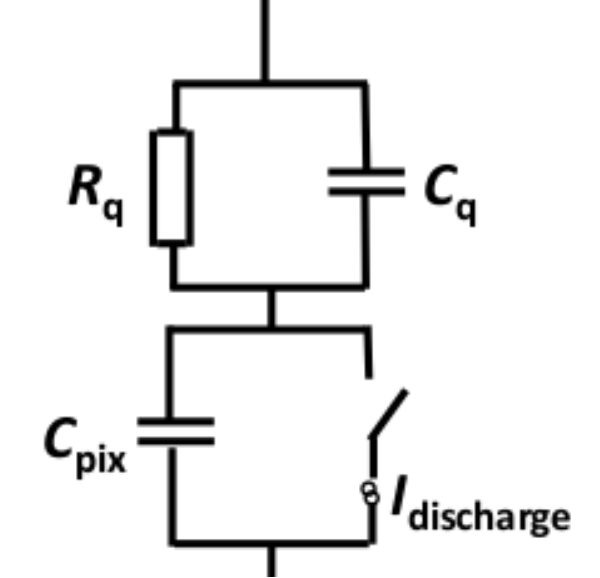}
    \caption{ }
   \end{subfigure}%
   \caption{(a) Schematic cross section, and (b) electrical model of a single SiPM pixel. }
  \label{fig:Fig1}
 \end{figure}

 The SiPMs investigated were produced by KETEK\,\cite{KETEK}.
 Four SiPMs of similar design and pixel sizes of 15$\times $15$\,\upmu \rm m^2$ to 100$\times $100\,$\upmu \rm m^2$ have been investigated.
 In addition, measurements were made on a single-pixel with an area of $1\,\rm mm^2$.
 Fig.\,\ref{fig:Fig1} shows a schematic cross section of a single pixel, and the electrical model used  for the analysis.
 Table\,\ref{tab:Table1} shows electrical parameters of the different SiPMs: The pixel capacitance, $C_{pix}$, the quenching resistance, $R_q$, and the capacitance $C_q$, which is in parallel to $R_q$.
 They have been determined by capacitance-frequency measurements at a bias voltage 27\,V using the method discussed in Ref.\,\cite{Xu:2014}.
 From the $C-V$\,measurements the depth of the avalanche zone was estimated to be $\approx 0.9\,\upmu $m.
 We note that MP15 comes from a different production run than the other three SiPMs.

\begin{table}  [!ht]
 \centering
 \footnotesize
  \begin{tabular}{c|c|c|c|c|c}
             & MP15 & PM25 & PM50 & PM100 & PNCV \\
   \hline
   \hline
   $N_{pix}$ & 4382 & 1600 & 400 & 100 & 1 \\
   \hline
   pitch [$\upmu $m] & 15 & 25 & 50 & 100 & 1000 \\
   \hline
   $C_{pix}$ [fF] & 18 & 69 & 330 & 15000 & $11\cdot 10^6$  \\
   \hline
   $C_q$ [fF] & < 5 & < 10 & 25 & 155 & -- \\
   \hline
   $R_q$ [k$\Omega$] & 750 & 500 & 340 & 410 & 0.13 \\
   \hline
   \hline
 \end{tabular}
  \caption{Parameters of the SiPMs investigated.
   \label{tab:Table1}}
\end{table}

 \section{Measurements and analysis}
  \label{sect:Measurements}
 \subsection{Current-breakdown voltage $V_I$}
 \label{subsect:Current}
\begin{figure}[!ht]
   \centering
   \begin{subfigure}[a]{0.5\textwidth}
    \includegraphics[width=\textwidth]{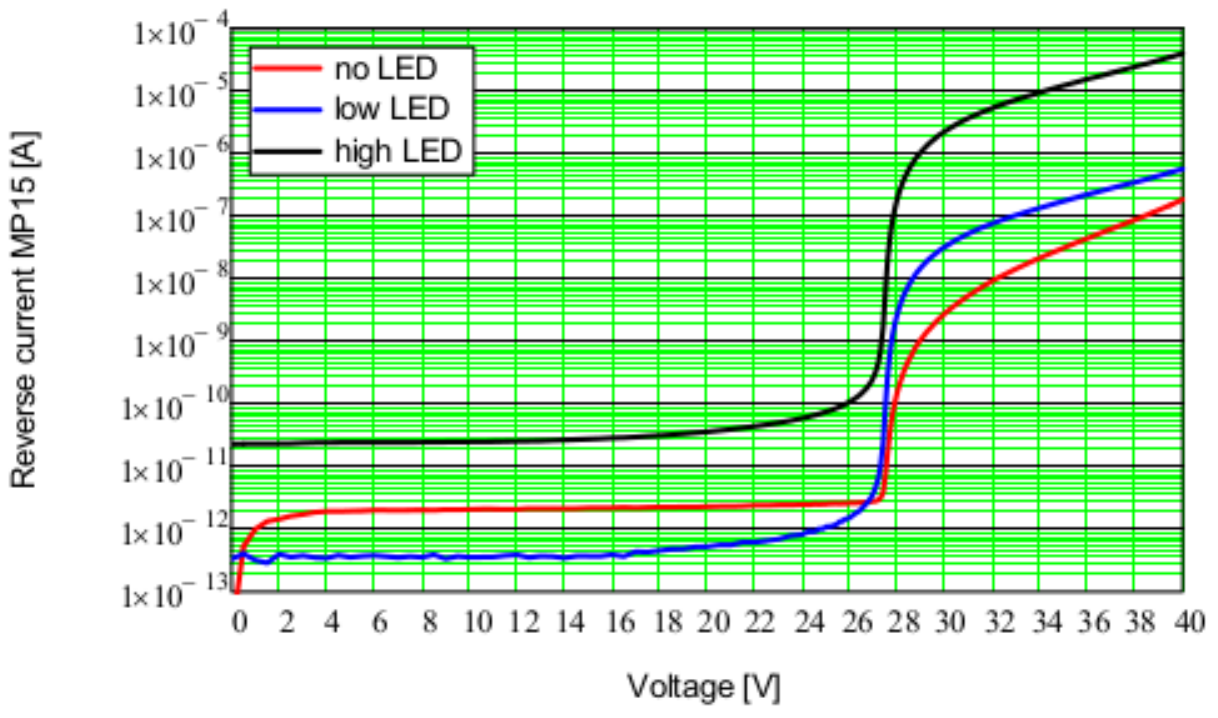}
    \caption{ }
   \end{subfigure}%
    ~
   \begin{subfigure}[a]{0.5\textwidth}
    \includegraphics[width=\textwidth]{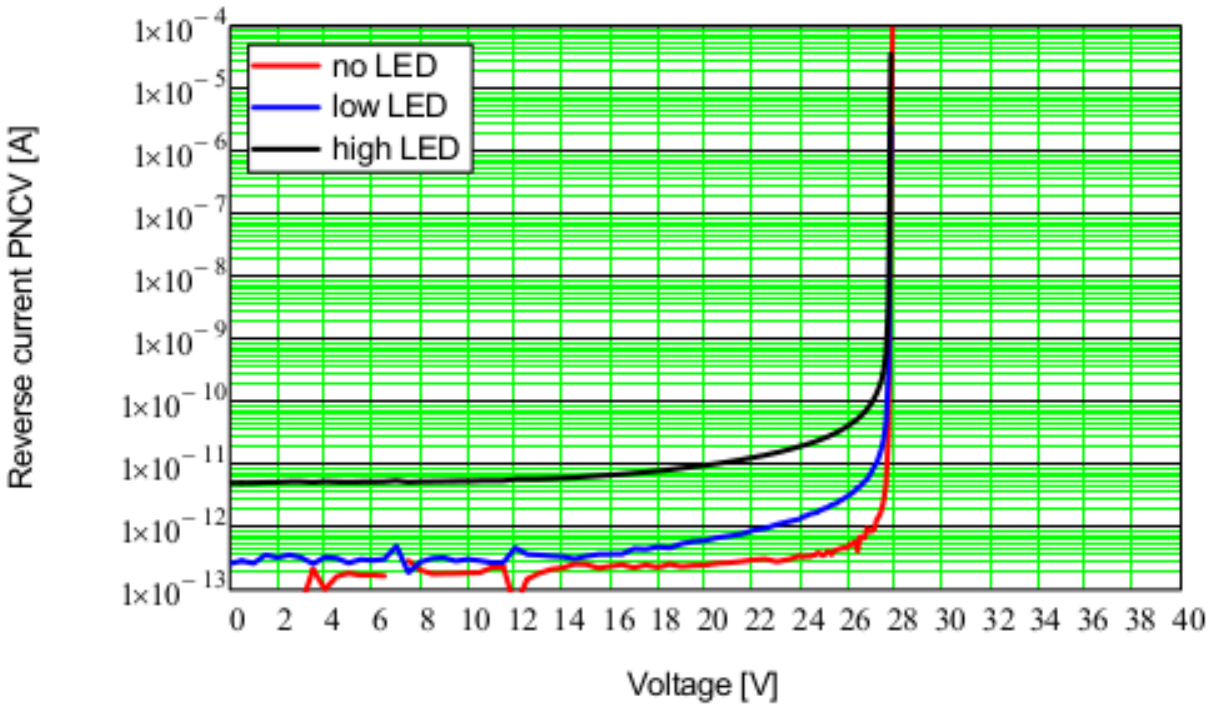}
    \caption{ }
   \end{subfigure}%
   \caption{(a) Current-voltage characteristics for MP15, and (b) for PNCV. Shown are the dark current, labeled \textit{no LED}, and the current for two light intensities \textit{low LED} and \textit{high LED}, with the \textit{no LED} current subtracted. }
  \label{fig:Fig2}
 \end{figure}

 For the determination of the \textit{current-breakdown voltage}, $V_I$, the SiPM currents as a function of the reverse voltage between 0 and 40\,V have been measured for the SiPM in the dark, and illuminated by an LED with a wavelength of 470\,nm.
 At this wavelength the light attenuation length at 20\,$^\circ $C is about $0.5\,\upmu$m, and most of the electron-hole pairs are generated close to the $p\, n$\,junction.
 The measurements were taken at $20\,^\circ$C without light, referred to as \textit{no LED}, and two light intensities, \textit{low LED} and \textit{high LED}.
 Fig.\,\ref{fig:Fig2} show as examples the measured $I-V$\,curves for MP15 and for PNCV.
 For the single-pixel PNCV, the value of $R_q = 130\,\Omega$ is too small to quench the avalanche, and at the breakdown voltage the current is limited only by the voltage source.
 The currents with LED are approximately constant up to about 10\,V, then show a gradual increase due to avalanche multiplication, which is followed by a sudden increase at the breakdown voltage.
 The dark current has a markedly different voltage dependence.
 For example, for  MP15 it is constant up to the breakdown voltage, and above the breakdown voltage the slope is steeper than the current due to the light.
 We interpret this as evidence that most of the sub-breakdown dark current misses the amplification region.
 The surface generation current from the Si-SiO$_2$ interface could be the main source of dark current below breakdown.

\begin{figure}[!ht]
   \centering
    \includegraphics[width=0.6\textwidth]{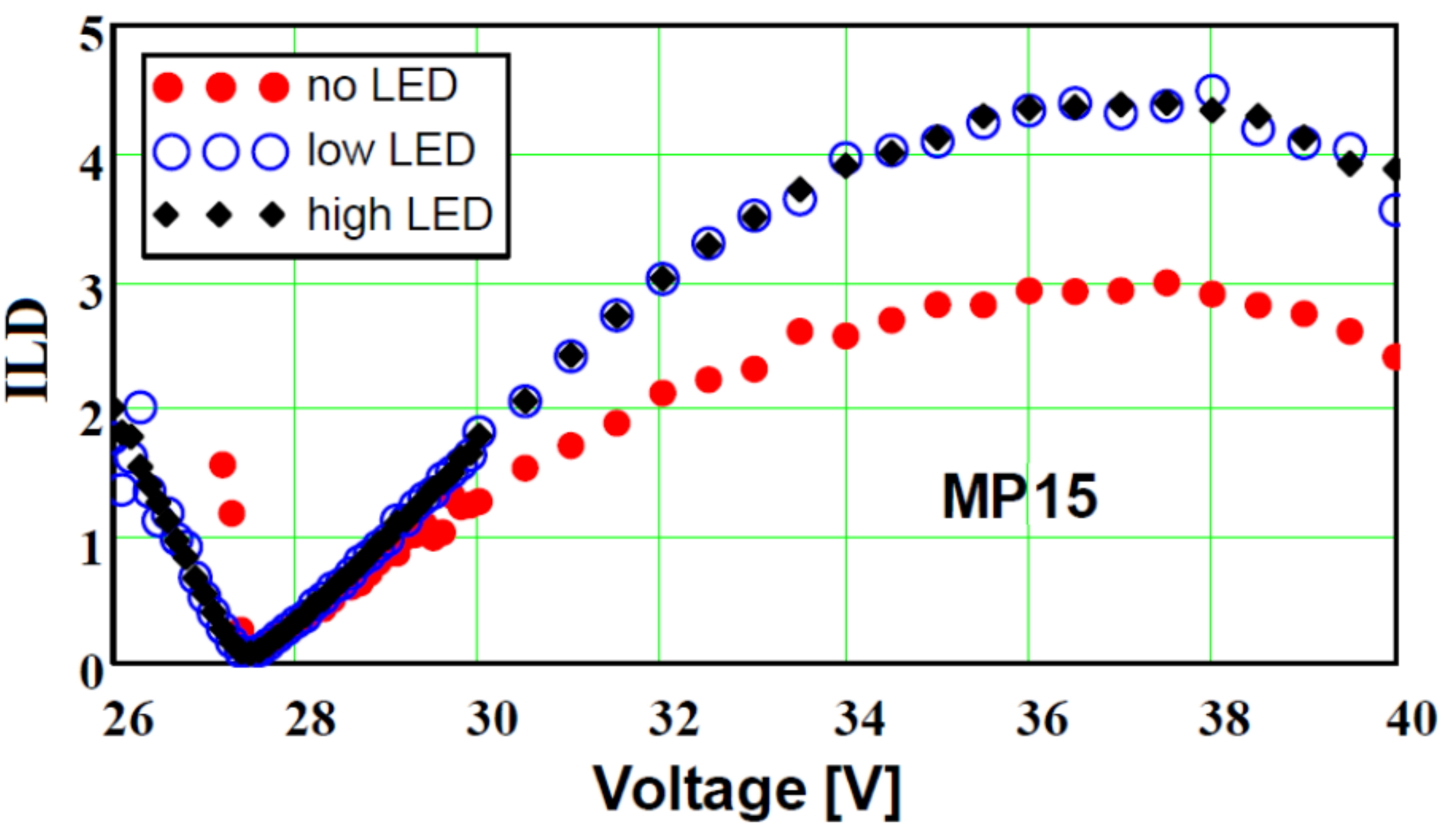}
   \caption{Inverse Logarithmic Derivative, $ILD$, for MP15. }
  \label{fig:Fig3}
 \end{figure}

 For the determination of $V_I$ from the $I-V$ measurement, the method of the Inverse Logarithmic Derivative, $ILD = ($d\,$\ln |I|/$d\,$V)^{-1}$, proposed in Ref. \cite{ILD:2014}, is used.
 The $ILD$ allows to characterize $I(V)$ curves:
 \begin{itemize}
   \item For a forward biased diode $I(V) \propto e^{\,q_0\,V/n\, kT}
       \rightarrow ILD = (n \cdot kT)/q_0  = const.$,
   \item for a reverse biased diode  below the breakdown voltage, $V_{bd}$, $I(V) \propto 1/|V-V_{bd}|^\alpha \rightarrow ILD = -|V-V_{bd}|/\alpha$, and
   \item for a SiPM with gain $G \propto (V-V_{bd})$, photon-detection efficiency $PDE \propto (V-V_{bd})^\kappa$, and cross talk $XT$, $I(V) \propto G \times PDE \times (1 +XT)$ for $V > V_{bd}$, $ILD \approx (V-V_{bd})/(1 + \kappa)$, as long as the contribution from $XT$ can be ignored.
 \end{itemize}
 We calculate $ILD$ using a cubic spline interpolation.
 Fig.\,\ref{fig:Fig3} shows $ILD$ for the $I-V$ curves of MP15 shown in Fig.\,\ref{fig:Fig2}.
 Below the breakdown voltage, $ILD$ shows a linear voltage dependence with a negative slope, and above breakdown, $ILD$ starts with a linear rise changing to a quadratic dependence at higher voltages.
 As expected, the values for \textit{low LED} and \textit{high LED} are compatible, whereas the values for \textit{no LED} above breakdown are smaller, reflecting the steeper slope of the $I(V)$ curve.
 We note that the positions of the minima of the $ILD$\,curves with and without illumination are the same.
 
 
 \begin{figure}[!ht]
   \centering
   \begin{subfigure}[a]{0.5\textwidth}
    \includegraphics[width=\textwidth]{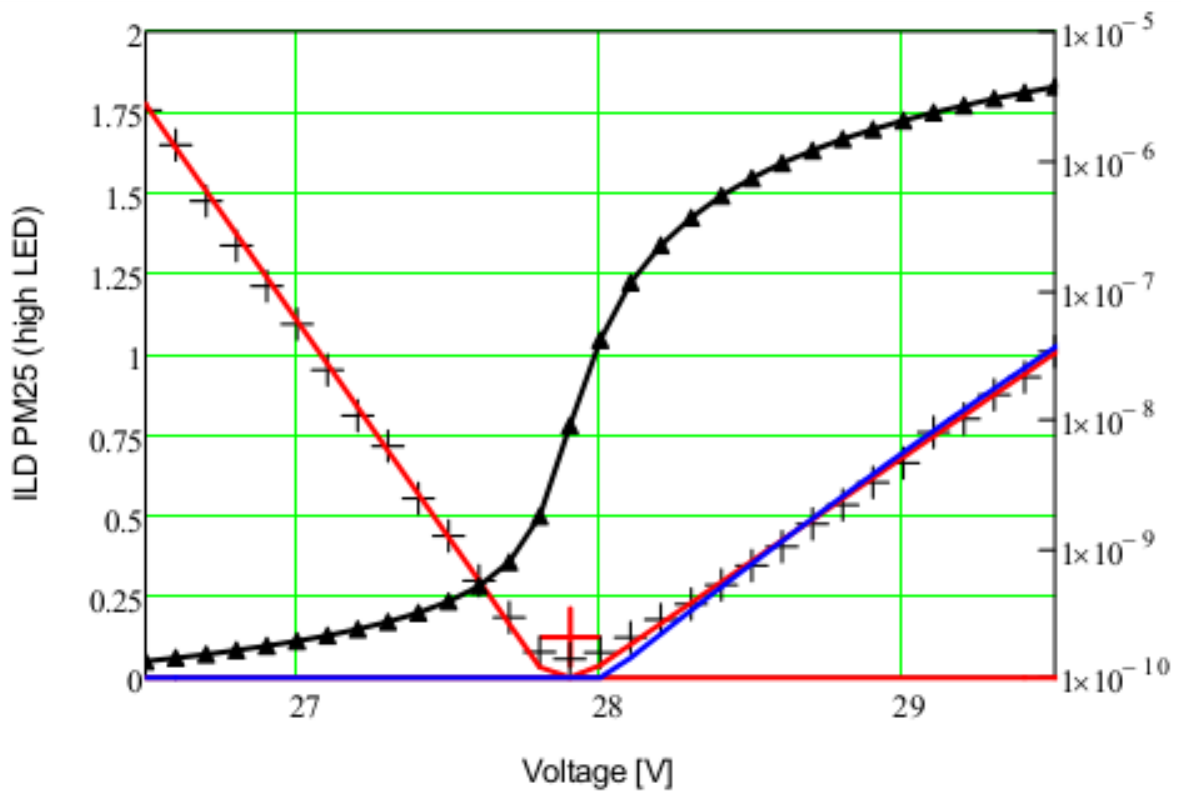}
    \caption{ }
   \end{subfigure}%
    ~
   \begin{subfigure}[a]{0.5\textwidth}
    \includegraphics[width=\textwidth]{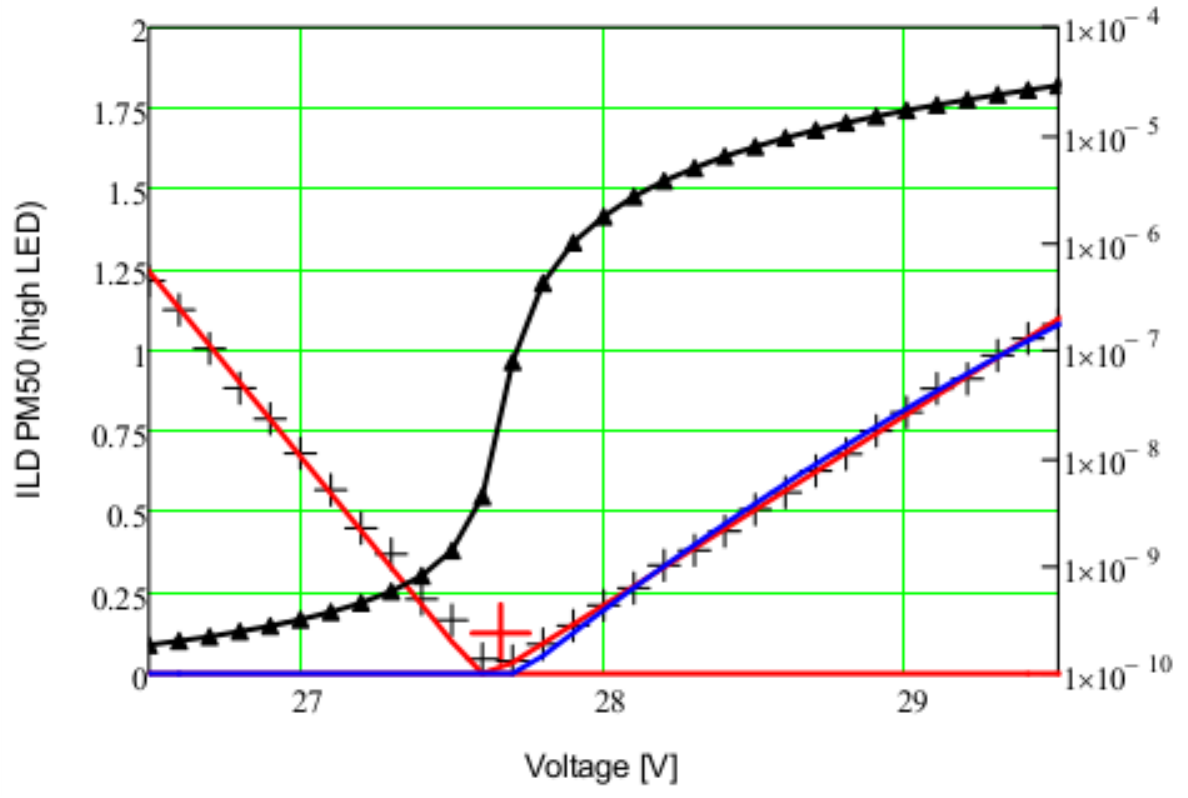}
    \caption{ }
   \end{subfigure}%
     
   \begin{subfigure}[a]{0.5\textwidth}
    \includegraphics[width=\textwidth]{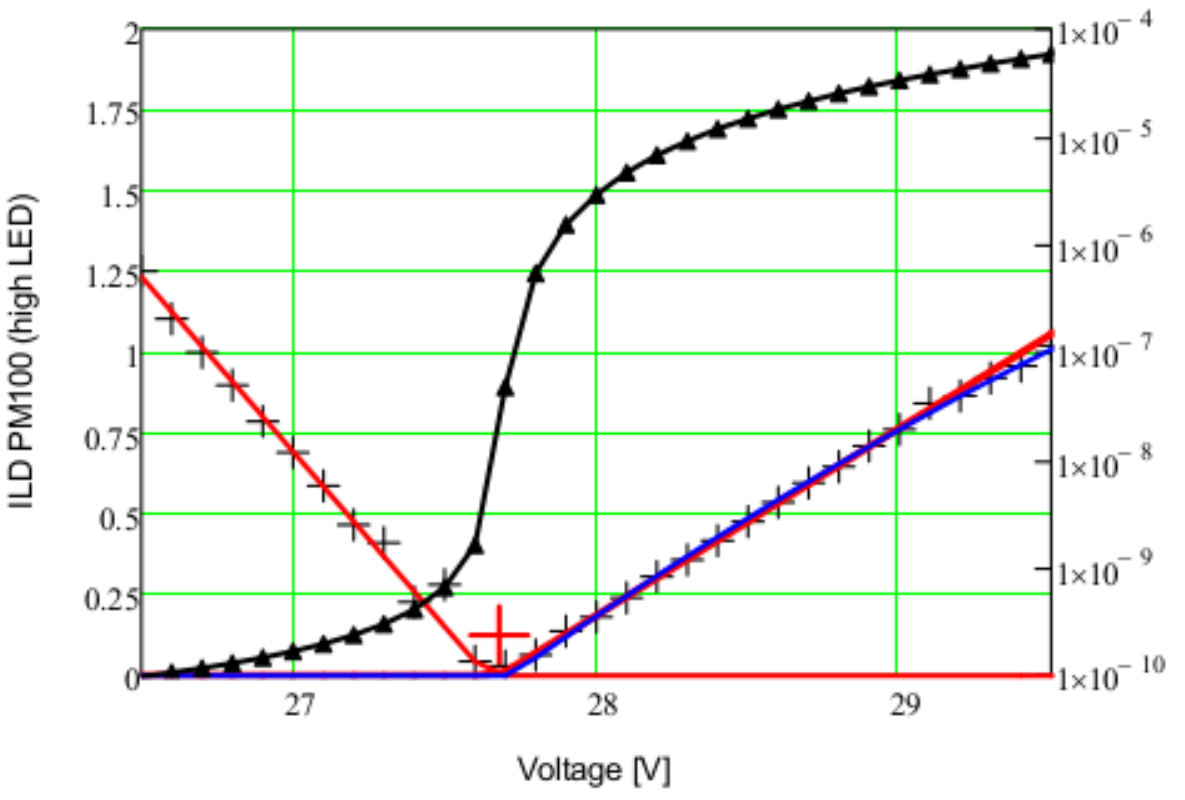}
    \caption{ }
   \end{subfigure}%
    ~
   \begin{subfigure}[a]{0.5\textwidth}
    \includegraphics[width=\textwidth]{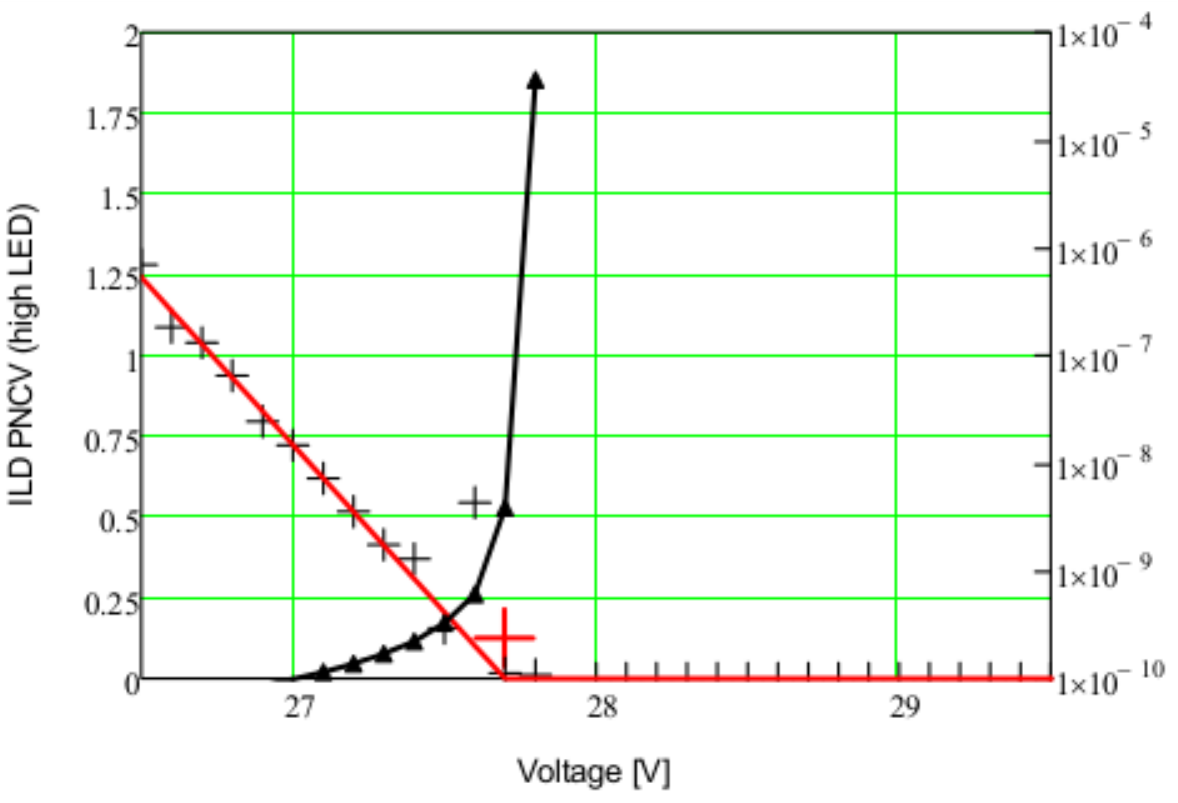}
    \caption{ }
   \end{subfigure}%
   \caption{Current (solid triangles, right scale) and $ILD$ (crosses, left scale) results for (a) PM25, (b) PM50, (c) PM100, and (d) PNCV for \textit{high LED}. The dashed, dotted lines are linear, quadratic fits to the $ILD$ values, respectively. The large crosses indicate the positions of the $ILD$\,minima. }
  \label{fig:Fig4}
 \end{figure}

 Fig.\,\ref{fig:Fig4} shows for PM25, PM50, PM100 and PNCV the $I - V$\,curves for \textit{high LED} and the corresponding $ILD$s.
 Four methods have been used to determine the current-breakdown voltage, $I_V$:
 The minimum of $ILD$ as obtained from a parabola of the three values around the minimal $ILD$\,values (red crosses in Fig.\,\ref{fig:Fig4});
 The intercepts with the $x$\,axis of straight-line fits to the values below and above breakdown (red lines);
 The intercept of a quadratic fit to the values above breakdown.
 The different values obtained for $V_I$ agree within $\pm 100$\,mV for MP15, and within $\pm 50$\,mV for PM25, PM50 and PM100.
 For the PNCV only the linear fit below breakdown is possible.
 The results of the fits including their spread are shown in Fig.\,\ref{fig:Fig7}, which summarizes the results from the different  breakdown-voltage determinations presented in the paper.
 We note that the values for $V_I$ obtained from the \textit{no LED} measurements are compatible.
 However, their uncertainties are larger.

 \subsection{Gain-breakdown voltage $V_G$}
  \label{subsect:Gain}
\begin{figure}[!ht]
   \centering
    \includegraphics[width=0.6\textwidth]{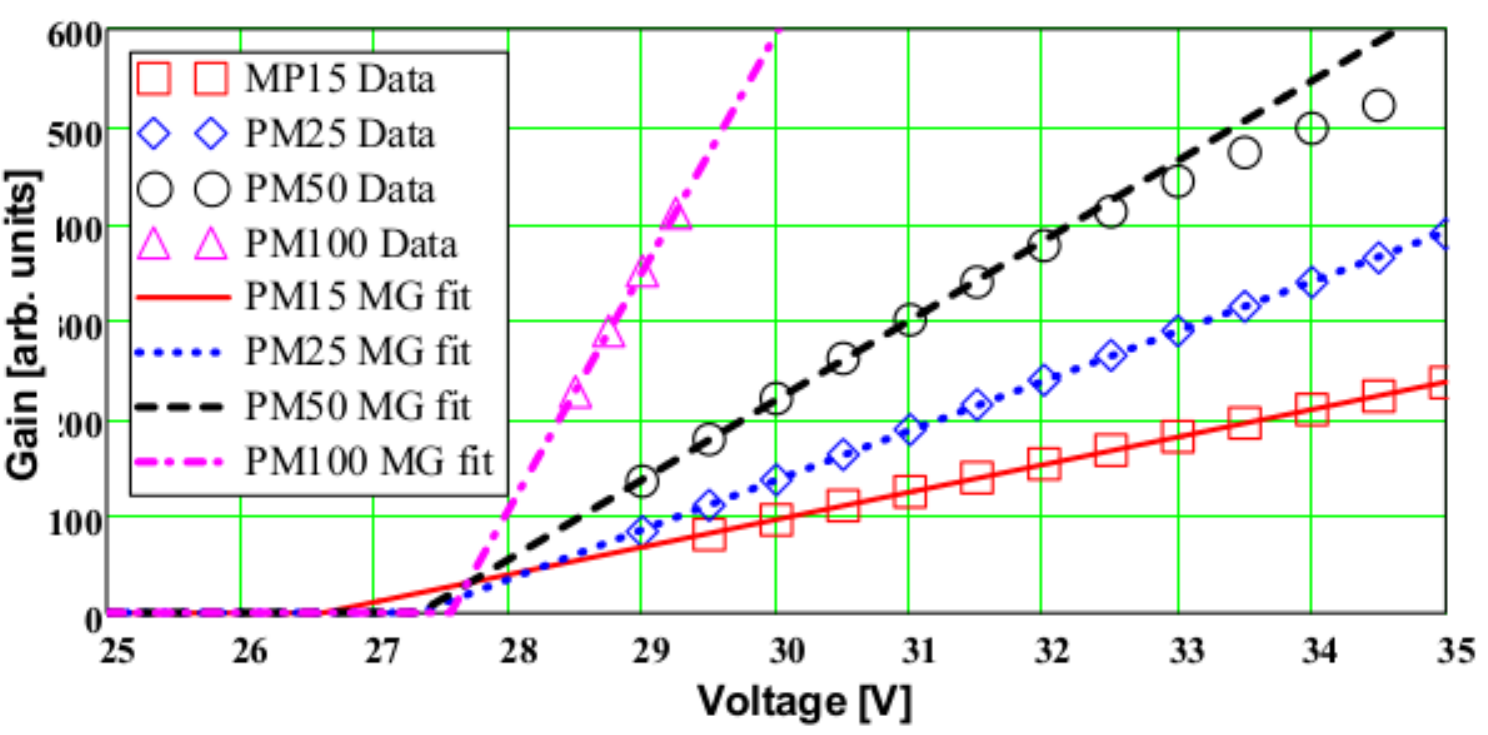}
   \caption{Gain versus bias voltage and straight line fits to determine the gain-breakdown voltage $V_G$. }
  \label{fig:Fig5}
 \end{figure}

 For the gain measurement, the signals from the SiPMs for \textit{no LED} and \textit{low LED} have been recorded using a CAEN QDC\,\cite{QDC} for several voltages above the breakdown voltage.
 A gate width of 100\,ns was used for the MP15 measurement, and 200\,ns for PM25, PM50 and PM100.
 The gains have been determined from the distances between the 0, 1, 2, etc. photo-electron $(pe)$ peaks.
 Various methods, like fits of separate Gauss functions to the $pe$ peaks (\textit{MG fit}) and Fourier transforms gave compatible results.
 For PM100, the dark-count rate was very high and for voltages above 29.5\,V a reliable extraction of the gain was not possible.
 Fig.\,\ref{fig:Fig5} shows the gain values as function of biasing voltage from the MG fits.
 The intercept of the fit by a linear function with the $x$\,axis gives the gain-breakdown voltage, $V_G$.
 Typical statistical uncertainties are $10 - 20$\,mV, and the systematic errors are estimated to be 50\,mV.
 The $V_G$ values are shown in Fig.\,\ref{fig:Fig7}.

 \subsection{Photodetection-breakdown voltage $V_{PD}$ }
   \label{subsect:PDE}
\begin{figure}[!ht]
   \centering
    \includegraphics[width=0.6\textwidth]{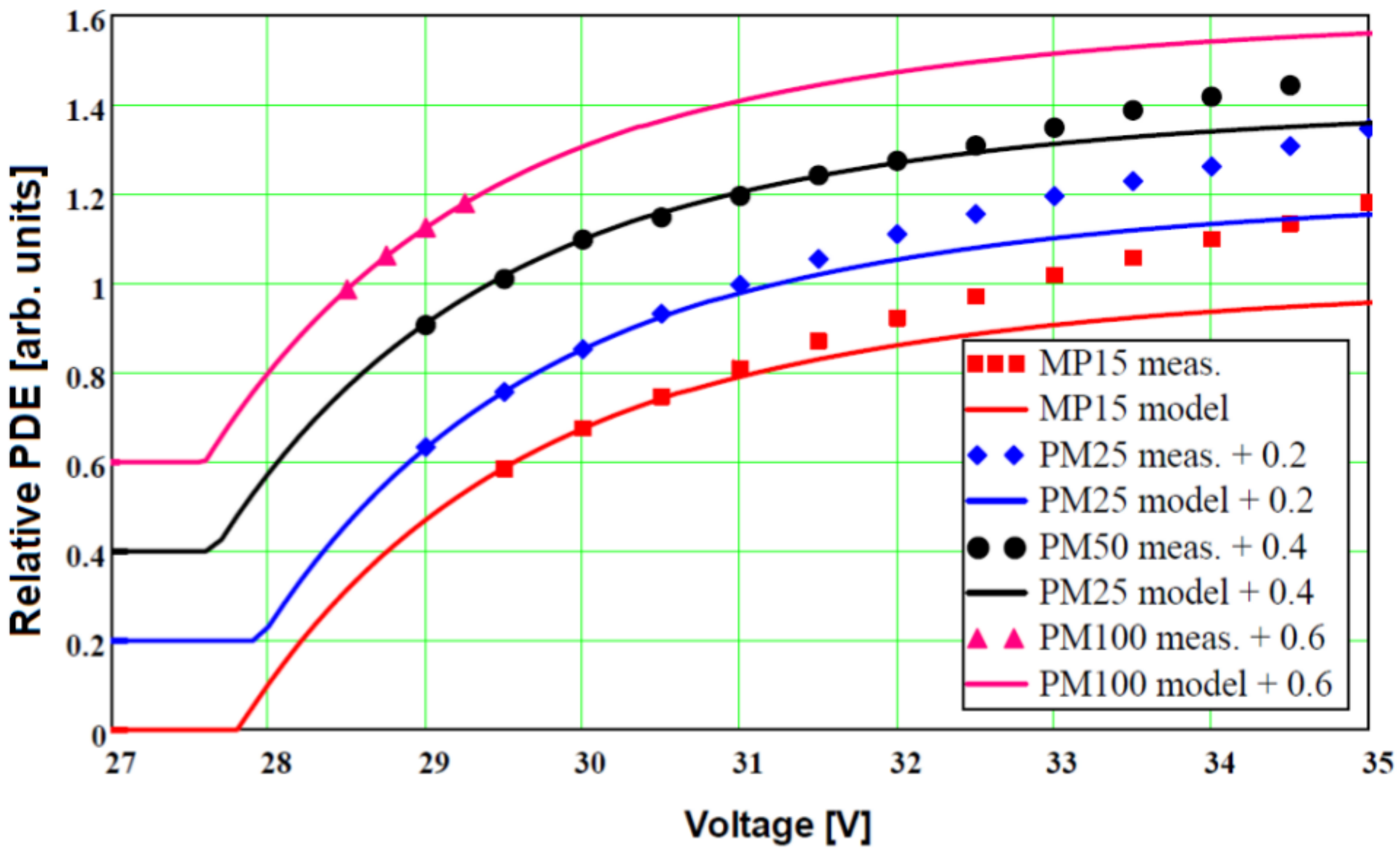}
   \caption{Relative $PDE$ as function of bias voltage, and comparison to the model calculations. The individual curves are shifted by multiples of 0.2. }
  \label{fig:Fig6}
 \end{figure}

 The \textit{low LED} data, used for the $V_G$ determination, have also been used for determining the relative Photon-Detection-Efficiency, $PDE$.
 The fraction of events with zero $pe$, $f_0$, has been obtained from the number events with a pulse area, $PA$, less than $1/2$ of the one-$pe$ $PA$, and the relative $PDE$ calculated using Poisson statistics: $PDE \propto -\ln (f_0)$.
 For the description of the voltage dependence of $PDE$, the following simplified model has been developed.
 A constant electric field $E$ has been assumed in an avalanche region of width $w$.
 The breakdown probability for an electron injected at the $p\,n$\,\,junction was calculated using the formulae given in Ref.\,\cite{McIntyre:1973} and the parameterizations of the ionization coefficients from Ref.\,\cite{Overstraten:1970}.
 The resulting values of $w$ has then been varied, until the calculated breakdown-probability curve matched the $PDE$\,data.
 The values of $w \approx 0.775\,\upmu$m and $E \approx 350$\,kV/cm are compatible with the results from the $C-V$ measurements.
 The  data$-$model comparisons are shown in Fig.\,\ref{fig:Fig6}.
 From the intercept of the model curve with the $x$\,axis, the values of $V_{PD}$, shown in Fig.\,\ref{fig:Fig7}, were obtained.
 We note that the voltage range for which the model  provides a good description of the data, increases with pixel size.
 This is expected, as pixel-edge effects are more significant for the smaller pixels.

 \subsection{Comparison of the breakdown-voltage results }
   \label{subsect:Comparison}

 
 \begin{figure}[!ht]
   \centering
   \begin{subfigure}[a]{0.5\textwidth}
    \includegraphics[width=\textwidth]{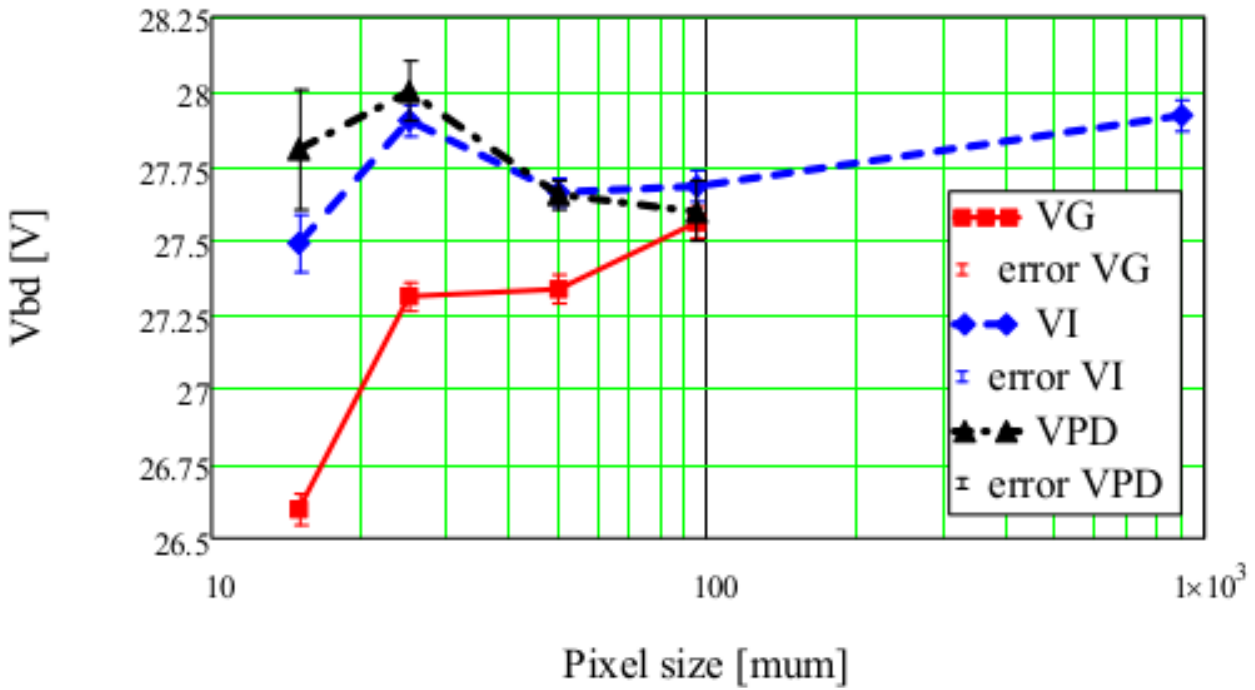}
    \caption{ }
   \end{subfigure}%
    ~
   \begin{subfigure}[a]{0.49\textwidth}
    \includegraphics[width=\textwidth]{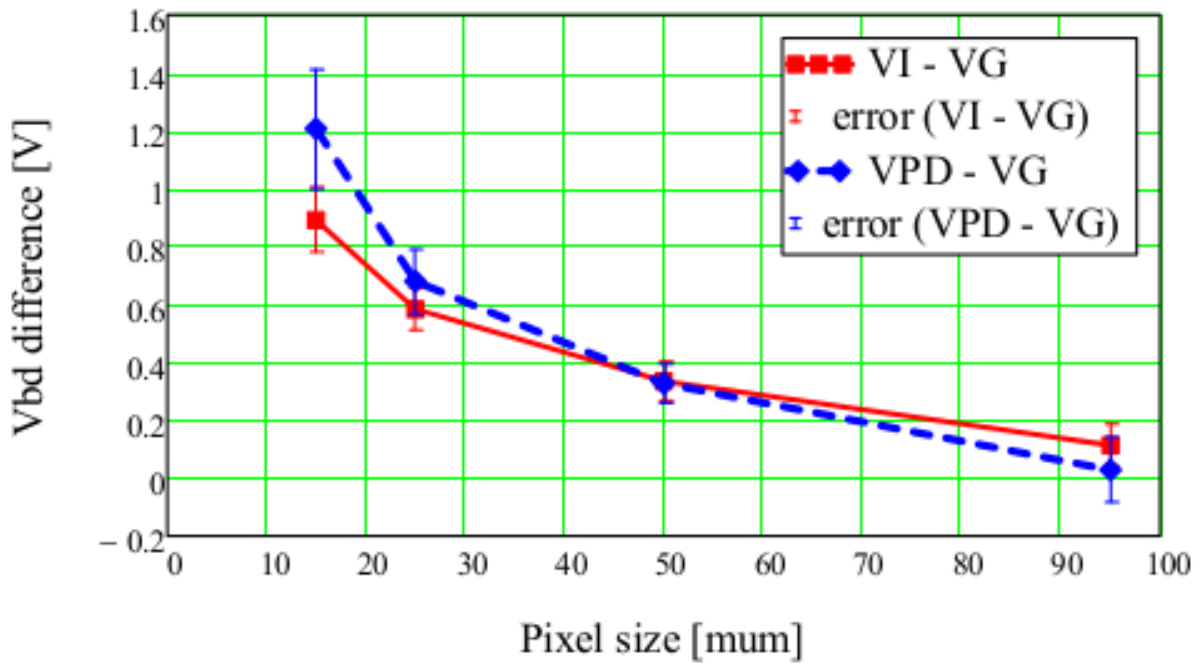}
    \caption{ }
   \end{subfigure}%
   \caption{(a) Values of the breakdown voltages $V_I$, $V_G$ and $V_{PD}$ as function of pixel pitch. (b) Differences $V_I - V_G$ and $V_{PD} - V_G$. }
  \label{fig:Fig7}
 \end{figure}
 
 Fig.\,\ref{fig:Fig7} compares the results of the three methods used to determine the breakdown voltages of the SiPMs as function of pixel pitch.
 The values of the current-breakdown voltages, $V_I$, and of the photodetection-breakdown voltages, $V_{PD}$, agree within their experimental uncertainties.
 With the help of the model calculation presented in Sect.\,\ref{subsect:PDE}, we identify them with $V_{turn\,on}$, the voltage at which the Geiger breakdown turns on.
 We note that within the uncertainties of the measurements, $V_{turn\,on}$ is independent of pixel pitch.
 The values of the gain-breakdown voltages $V_G$ show a significant decrease with decreasing pixel size:
 For the SiPM with 100\,$ \upmu $m pitch the difference $V_I - V_G$ is compatible with zero, whereas $V_I - V_G \approx 1$\,V for 15\,$ \upmu $m pitch.
 A lower value for $V_G$ than for $V_{turn\,on}$ is not unexpected, as the charge multiplication required to form a plasma channel and turn on a Geiger discharge is significantly larger, than the one required to  maintain the plasma channel\,\cite{Marinov:2007}.
 However, we do not understand, why the difference changes with pixel pitch.
\section{Summary and conclusions}
 \label{sect:Summary}
 This paper presents three methods to determine the breakdown voltages of SiPMs:
\begin{enumerate}
   \item $V_I$ from the current$-$voltage dependence,
   \item $V_G$ from the gain$-$voltage dependence,
   \item $V_{PD}$ from the photodetection efficiency$-$volt\-age dependence.
\end{enumerate}
 These methods were used to characterize SiPMs produced by KETEK with pixel pitches of 15, 25, 50 and 100\,$\upmu $m.
 In addition, for a single-pixel SiPM with 1\,mm pitch without quenching resistor, $V_I$ has been determined from  $I - V$\,measurements below breakdown.
 It is found that $V_I \approx V_{PD}$ and that for the SiPMs from one production run, $V_I$ is independent of pixel size.
 Using a model calculation for the probability of a Geiger discharge for an electron injected at the $p\,n$\,\,junction, $V_I$ and $V_{PD}$ are identified as the turn-on voltage of the Geiger discharge.
 It is observed that $V_G$ has a significant dependence on pixel size: $V_I - V_G$ is approximately 1\,V for the (15\,$\upmu $m)$^2$ pixel, decreasing to zero for (100\,$\upmu $m)$^2$.
 Whereas $V_I > V_G$ is expected, the dependence on pixel size is not understood.
 We conclude that the $I-V$\,measurement does not determine $V_G$, which is the relevant quantity for the user, as the SiPM gain at the bias voltage $V$ is proportional to the voltage $V - V_G$.
 The observed differences in breakdown voltages provides evidence that the turn-on and turn-off voltages of Geiger discharges are different, which should be taken into account when modeling the response of SiPMs.

\newpage
\section*{Acknowledgement}
 \label{sect:Acknowledgement}
 We would like to thank Florian Wiest and his colleagues from KETEK for providing the SiPMs samples and for fruitful discussions.

\section{List of References}

\label{sect:Bibliography}



\begin{thebibliography}{0}




 \bibitem{KETEK}
  KETEK, Hofer Str. 3, D-81737 Munich, Germany
   \url{http://www.ketek.net}.

  \bibitem{Xu:2014}
 Ch.\,Xu et al.,
   Nuclear Instruments and Methods in Physics Research Section A~762 (2014) 149--161,
   doi:10.1016/j.nima.2014.05.112.

  \bibitem{ILD:2014}
 E.\,Garutti, et al.,
 Technology and Instrumentation in Particle Physics 2014, 2-6 June 2014, Amsterdam,
 PoS(TIPP2014)070.

 \bibitem{QDC}
   CAEN qS srl, Via Vetraia 11, 55049 - Viareggio (LU) - Italy,
   \url{http://www.caen.it}.

  \bibitem{McIntyre:1973}
   R.\,J.\,McIntyre, IEEE-Transactions of Electronic Devices 20(7), July 1973, 637--641.

 \bibitem{Overstraten:1970}
  R.\,van\,Overstaeten and H.\,de\,Man,
   Solid-State Electronics vol.\,13 (1), 1970, 583--608.

 \bibitem{Marinov:2007}
  O.\,Marinov, J.\,Dean and J.\,A.\,Jimenez Tejada
   Journal of Applied Physics 101, 064515 (2007),
    doi:10.1063/1.2654973.

\end{thebibliography}


\end{document}